\documentclass[12pt,preprint]{aastex}
\begin{document}

\title{The infrared counterpart to the magnetar 1RXS~J170849.0$-$400910}
\author{Martin Durant and Marten H. van Kerkwijk}
\affil{Department of Astronomy and Astrophysics, University of
  Toronto\\  60 St. George St, Toronto, ON\\ M5S 3H8, Canada }

\begin{abstract}
We have analyzed both archival and new infrared imaging observations
of the field of the Anomalous X-ray Pulsar 1RXS~J170849.0$-$400910, in
search of the infrared counterpart. This field has been previously
investigated, and one of the sources consistent with the position of
the AXP suggested as the counterpart. We, however, find that this
object is more likely a background star, while another object within
the positional error circle has non-stellar colors and shows evidence
for variability. These two pieces of evidence, along with a
consistency argument for the X-ray-to-infrared flux ratio, point to
the second source being the more likely infrared counterpart to the
AXP.
\end{abstract}
\maketitle

\section{Introduction}
The anomalous X-ray pulsars (AXPs) are a class of neutron stars,
numbering about half a dozen, which are radio-quiet, with periods of
the order $\sim10$\,s and estimated ages of $10^3$ to $10^5$\,yr. Like the soft
gamma-ray repeaters, they
are thought to be {\em magnetars}, whose emission is powered by the decay
of a super-strong magnetic field ($\sim 10^{15}$\,G). See Woods \&
Thompson (2004) for a review of the known magnetars and their
properties. 

While energetically, the emission at X-ray energies dominates, optical
and infrared photometry of AXPs is giving interesting constraints on the
physical processes of the stellar magnetospheres and
environment. Recently, Wang et 
al. (2006) identified a mid-infrared and K-band excess around a 
magnetar, 4U~0142+61, which they interpret as thermal emission from a
passively illuminate dusty fall-back disc. It would be interesting to
see whether this is a generic property of the AXPs. If so, it might
explain the consistency of K-band to soft X-ray flux ratios for most
of the AXPs (Durant \& van Kerkwijk, 2005a).

1RXS~J170849.0$-$400910 is a magnetar with 11s pulsations, discovered
in the soft X-ray band by {\em ROSAT} and {\em ASCA} (Sugizaki et
al. 1997). Recently, a hard X-ray component ($\sim100$\,keV) to its
spectrum has been found, which dominates the magnetar energetics 
(Kuiper at al. 2006).

Israel et al. (2003) reported a tentative identification of the
infrared counterpart to 1RXS~J170849.0$-$400910, based on
near-infrared H- and K-band adaptive optics observations with the
Adaptive Optics Bonette (AOB) on the Canada-France-Hawaii Telescope
(CFHT), and further JHK photometry from the European Southern
Observatory's New Technology Telescope (ESO NTT). They found two
possible faint counterparts in the positional error circle, Stars
``A'' and ``B'' separated by only 0\farcs26 (see images below). Israel
et al.\, suggested Star ``A'' was the more likely counterpart, based
on its peculiar colors. Below, we present a re-analysis of their CFHT
data, together with our own data and deep archival Very Large
Tetescope (VLT) imaging. We first describe these datasets and our
analysis methods, followed by the lines of argument which lead to our
conclusion that in fact the true counterpart is Star ``B''.

\section{Observation and Analysis}
We analyzed observations made with Magellan/PANIC and archival
observations from CFHT/AOB and VLT/NACO (see Table \ref{obs} and below
for
details). The Magellan observations provide the widest field of view,
and a uniform PSF and background; they are therefore the best images
to base our photometric calibration on.  The CFHT and VLT observations both made
use of adaptive optics (AO) in order to reduce the size of stellar PSFs
and thus increase the signal to noise ratio as well as reduce the
problem of blending. Unfortunately, this comes at the cost of a PSF
which has a complicated shape and varies with position on the image
(particularly for shorter wavelengths). We thus calibrate the AO
images using Magellan as our baseline.

\begin{deluxetable}{lccccc}
\tablecaption{List of observations. \label{obs}}
\tablewidth{0pt}
\tablehead{
\colhead{Date} & \colhead{Telescope} & \colhead{Instrument} &
\colhead{Filter} & \colhead{Exposure Time (s)} & \colhead{FWMH\tablenotemark{a}}}
\startdata
2002-08-17 & CFHT & AOB/KIR & K' & $60\times45$ & 0.14 \\
 &  & & H & $60\times45$ & 0.14 \\
2003-06-06 & Magellan Clay & PANIC & J & $60\times9$ & 0.44 \\
 & & & H & $60\times9$ & 0.35 \\
 & & & K$_S$ & $25\times21$ & 0.31 \\
2003-06-07 & Magellan Clay & PANIC & J & $60\times18$ & 0.38 \\
 & & & H & $60\times9$ & 0.4 \\
 & & & Y & $60\times9$ & 0.36 \\
 & & MagIC & I & $300\times15$ & 0.44 \\
2003-06-20 & VLT-UT3 & NA-CO & J & $60\times40$ & 0.11 \\
 & & & H & $10\times40$ & 0.10 \\
 & & & K$_S$ & $20\times80$ & 0.08
\enddata
\tablenotetext{a}{Typical stellar profile size in arcsec. For AO
  images, measured close to the stars on interest.}
\end{deluxetable}

\begin{figure}
\includegraphics[width=0.32\hsize,angle=270]{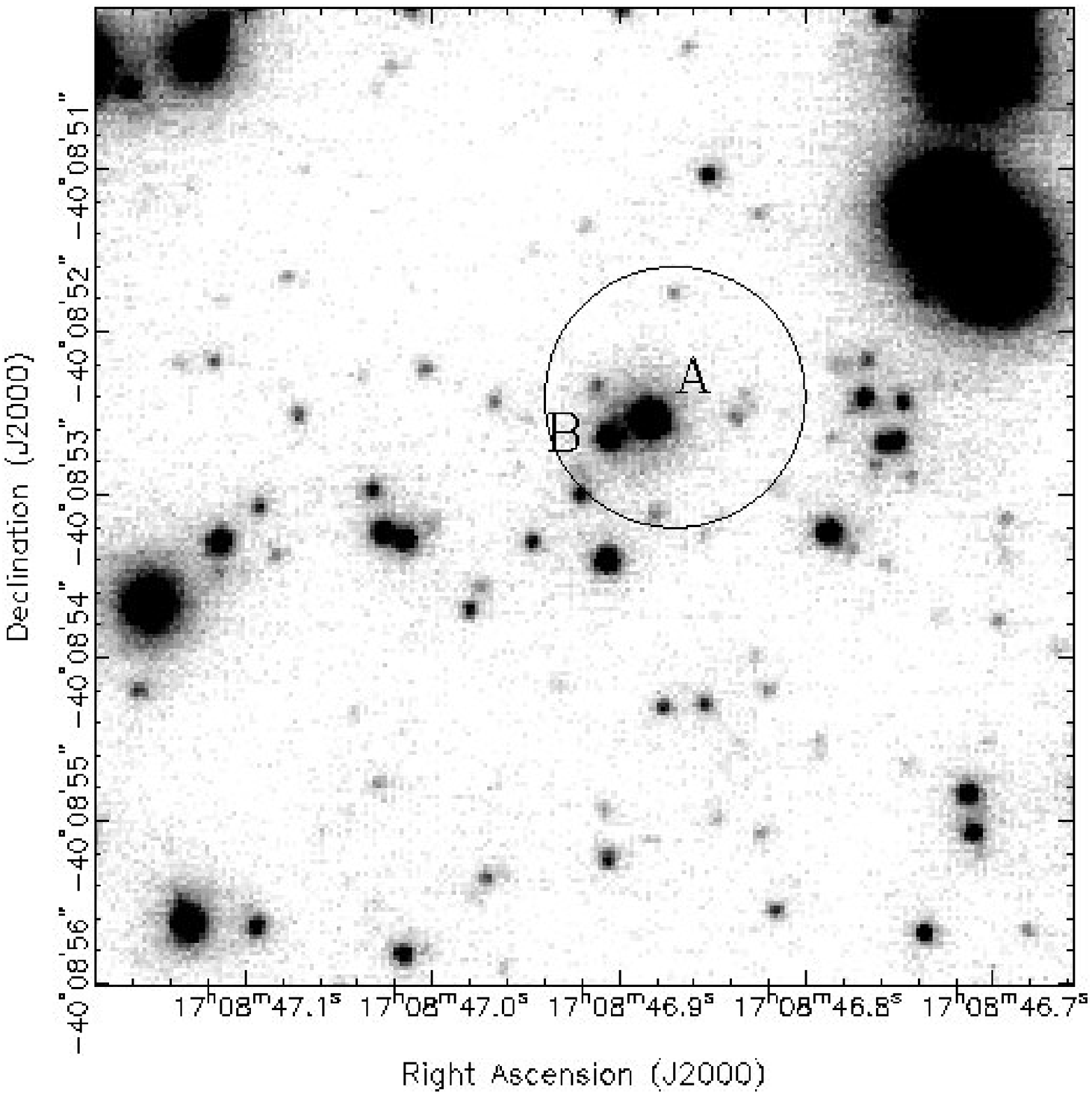}
\includegraphics[width=0.32\hsize,angle=270]{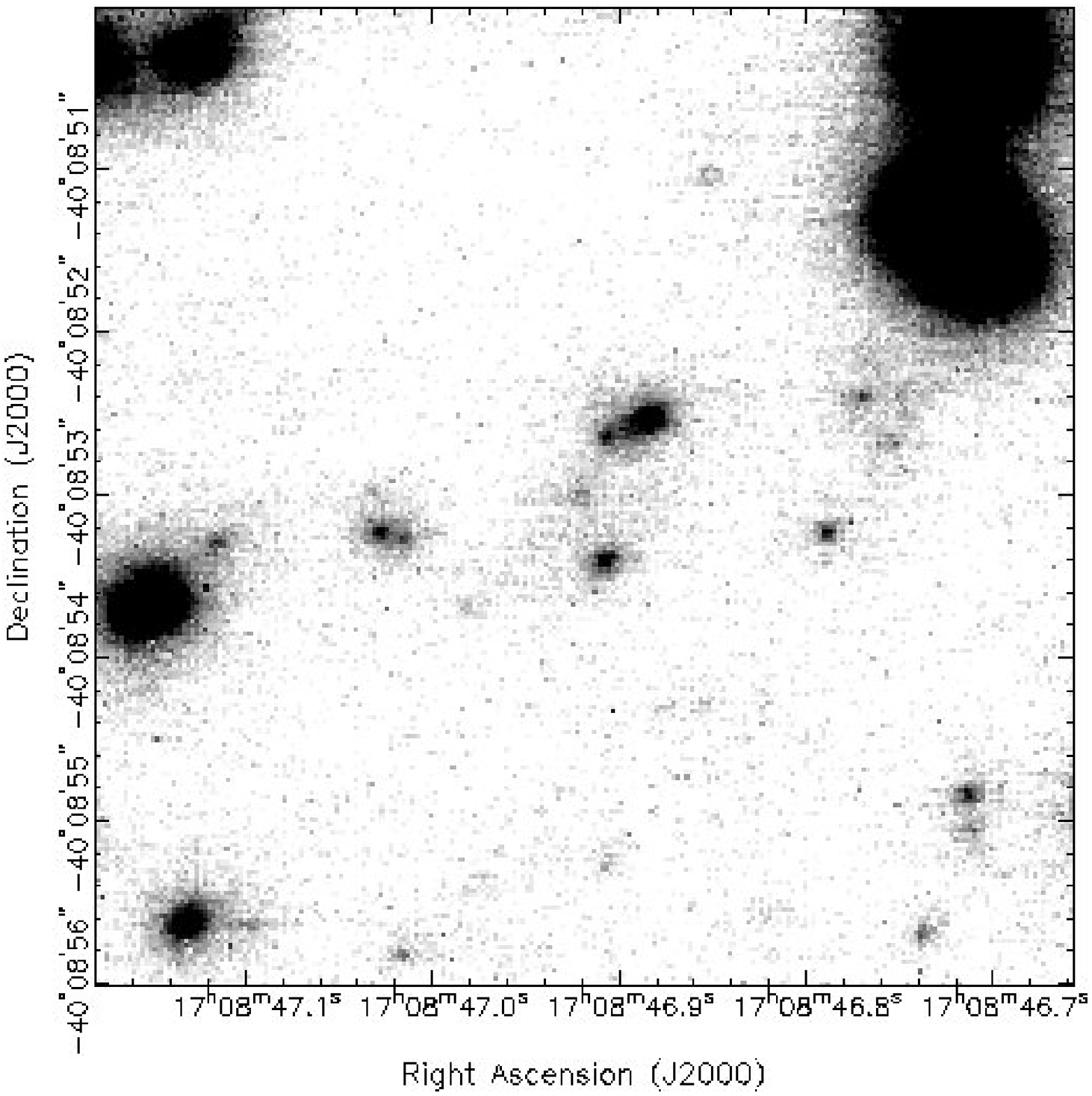}
\includegraphics[width=0.32\hsize,angle=270]{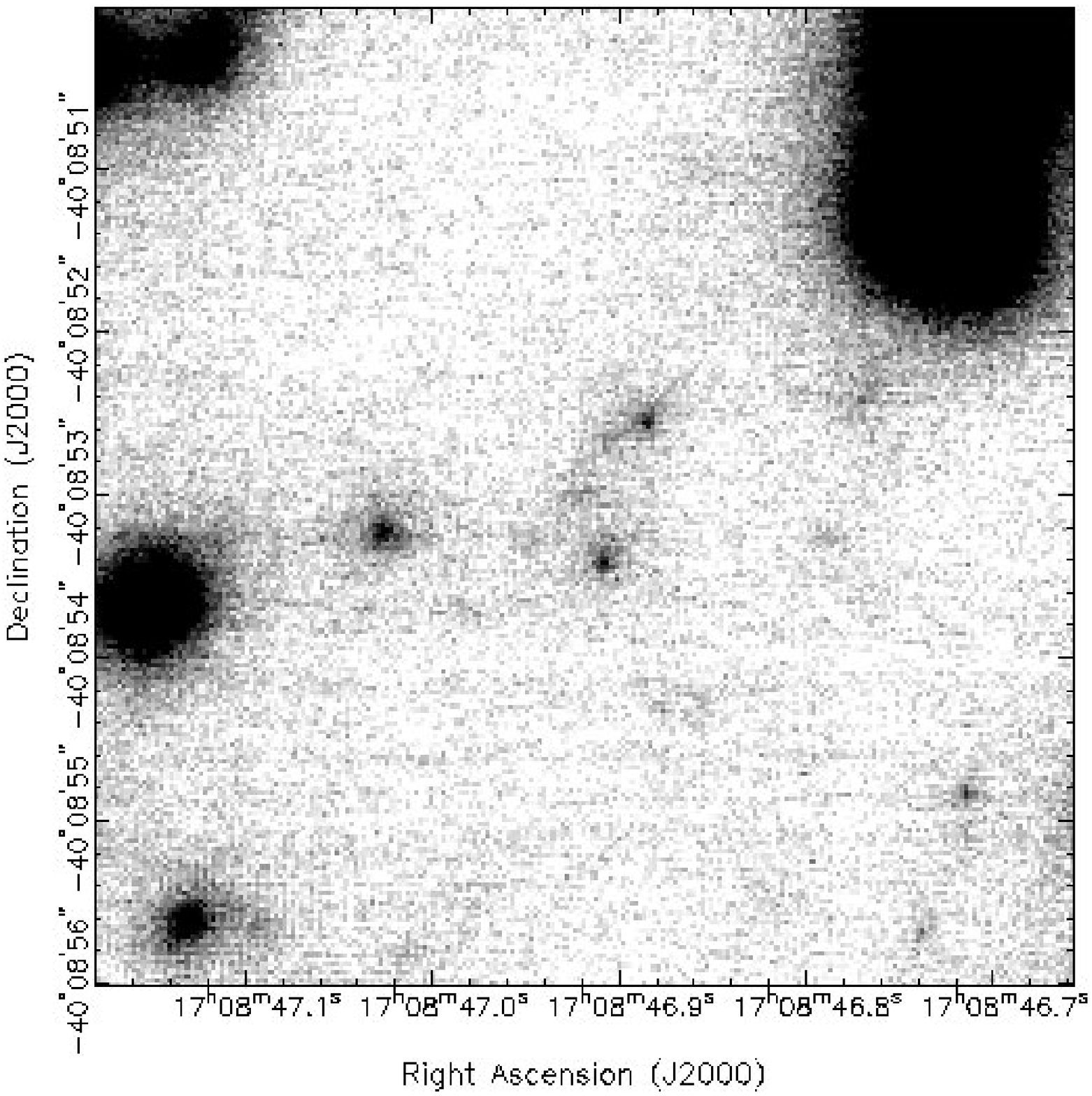}\\
\includegraphics[width=0.32\hsize,angle=270]{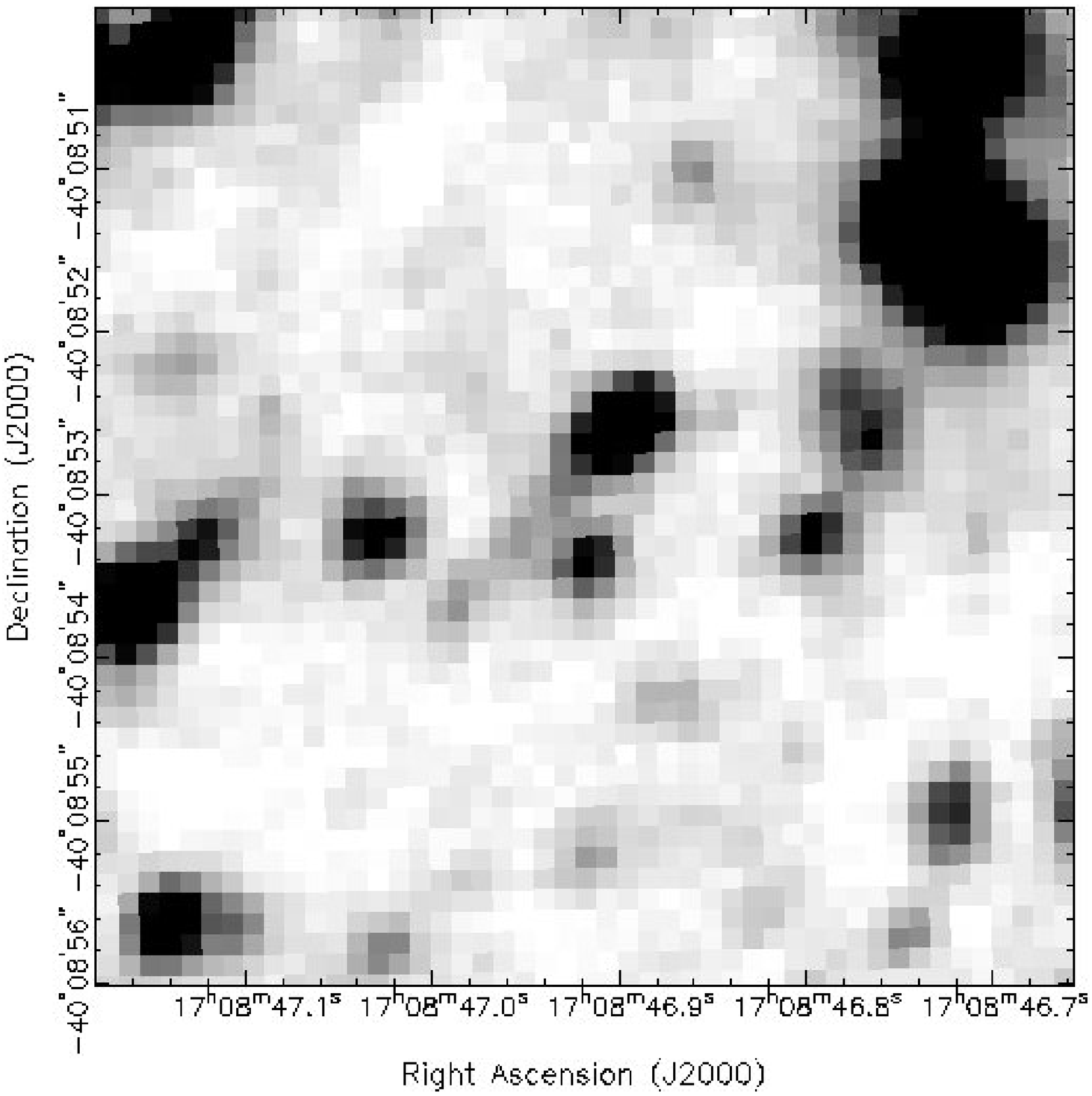}
\includegraphics[width=0.32\hsize,angle=270]{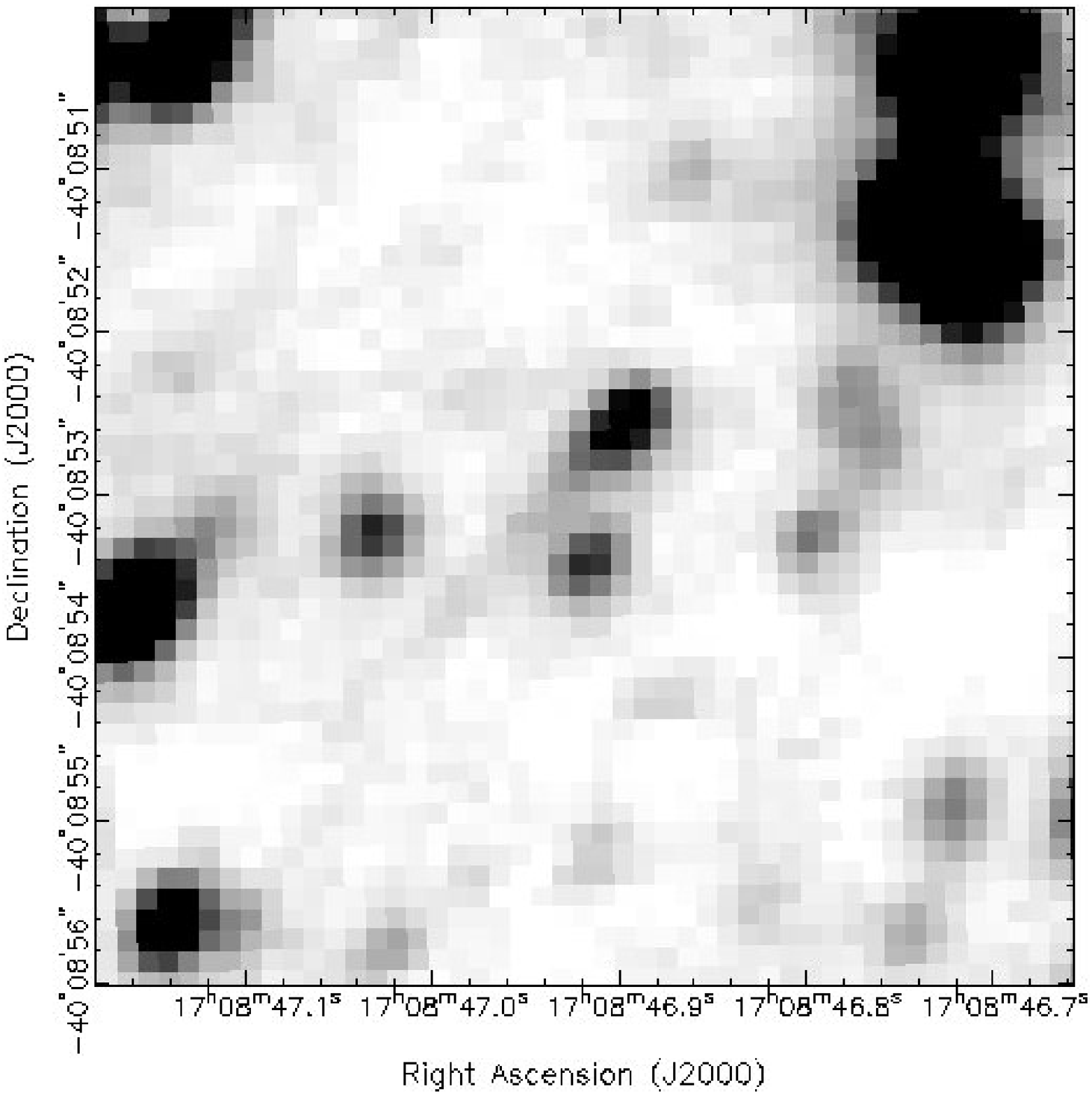}
\includegraphics[width=0.32\hsize,angle=270]{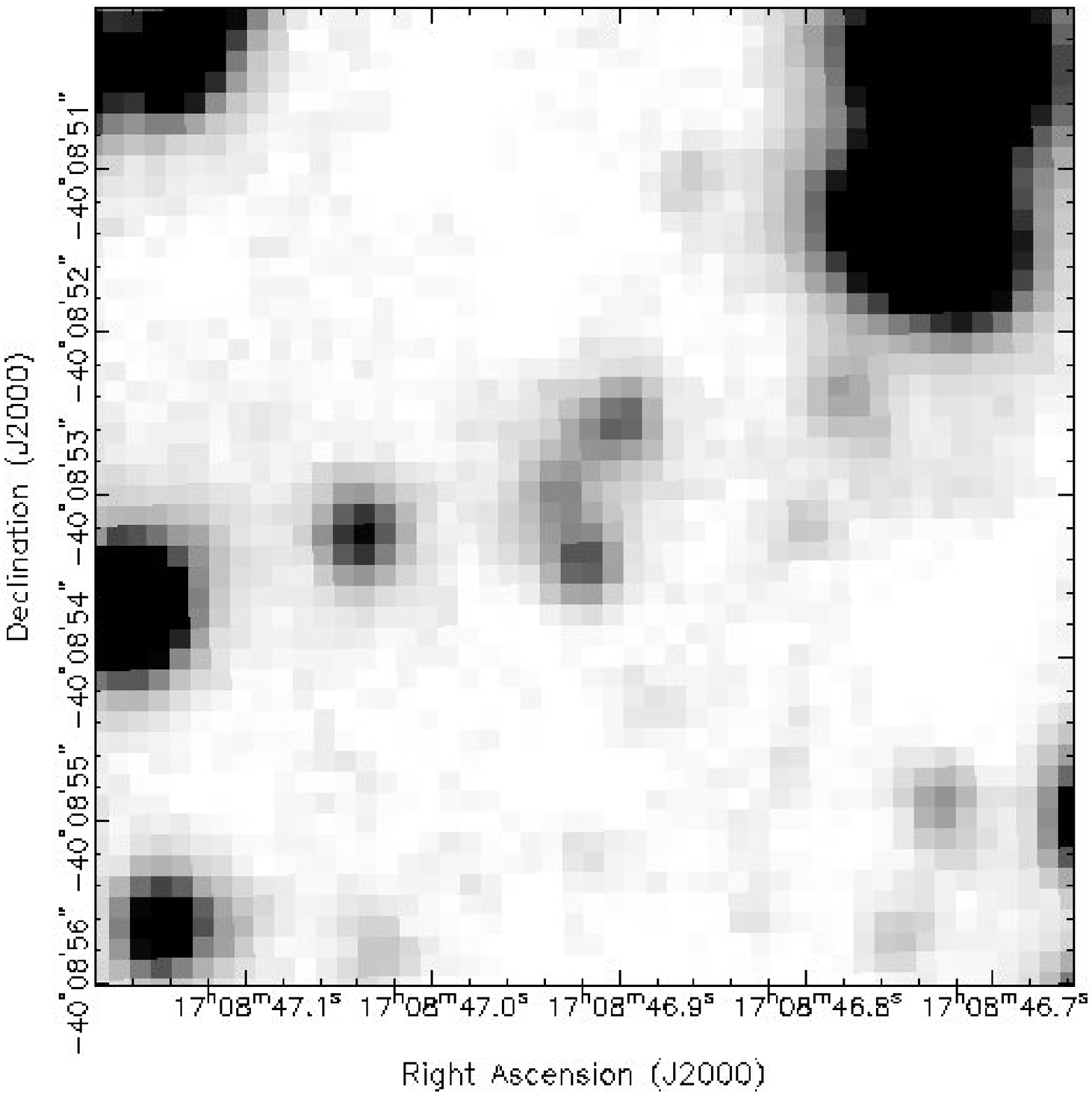}\\
\includegraphics[width=0.32\hsize,angle=270]{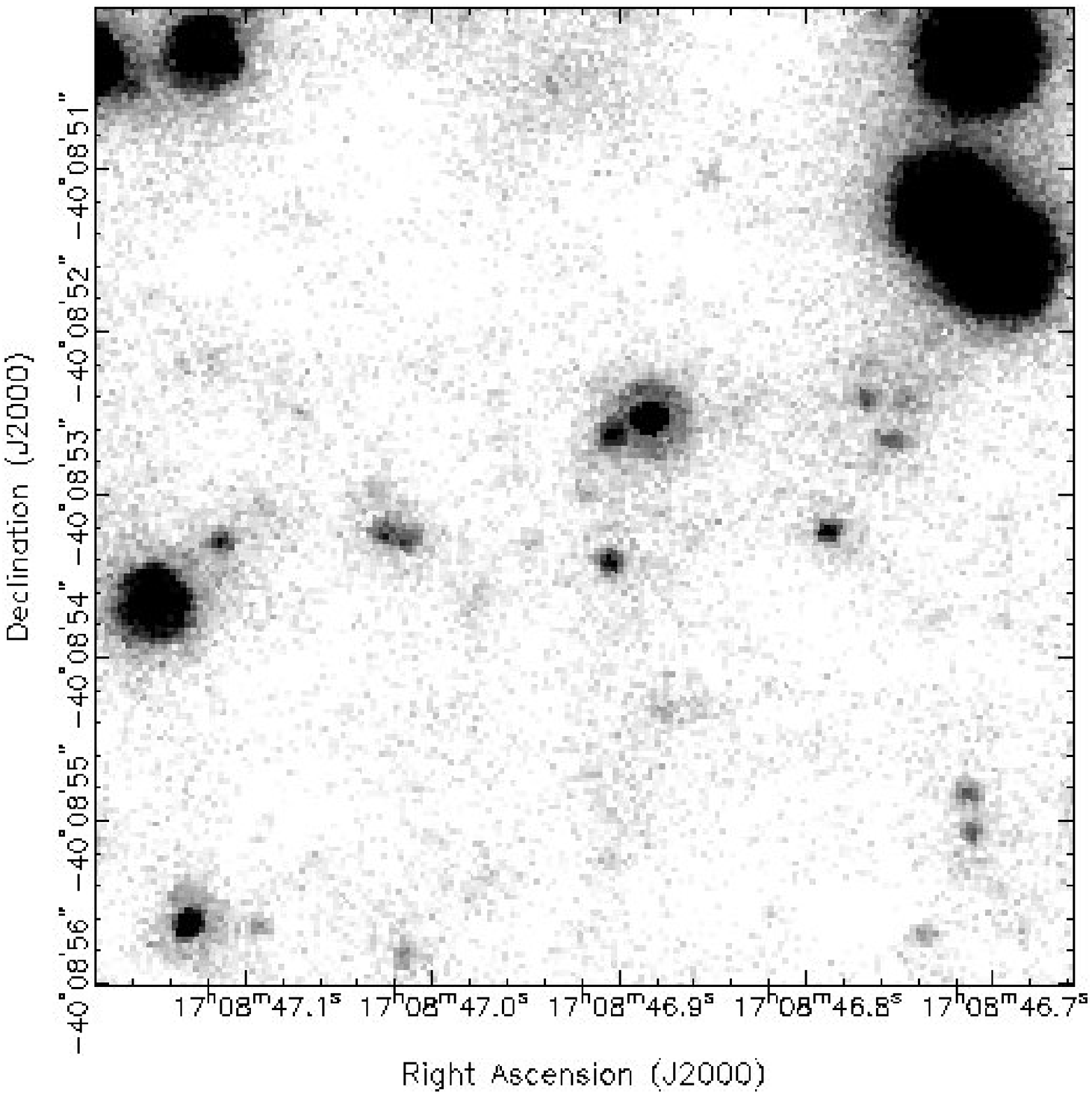}
\includegraphics[width=0.32\hsize,angle=270]{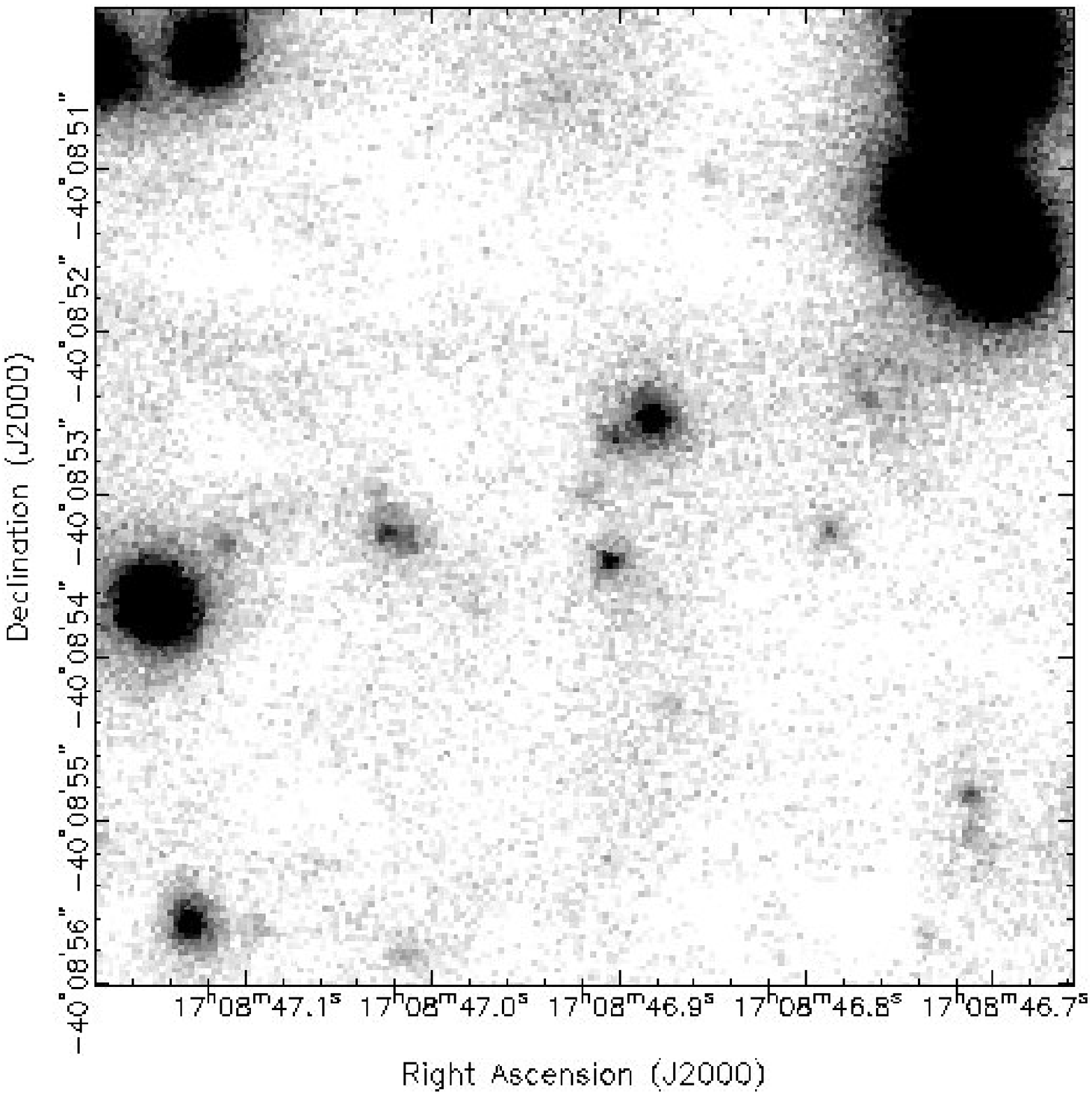}
\caption[0]{Images of the field of 1RXS~J170849.0$-$400910, from
  VLT/NACO (top), Magellan/PANIC (middle) and CFHT/AOB (bottom), with
  KHJ from left to right. In the VLT/NACO K-band image,
  the 90\%-confidence 0\farcs8-radius {\em Chandra} position error
  circle (at 90 \%) is shown and the two candidate counterparts, Stars
A and B are labeled.}\label{pics}
\end{figure}

\subsection{Magellan}
We imaged 1RXS~J170849.0$-$400910 in the K$_S$HJ bands using PANIC
(Persson's Auxilliary Nasmyth Infrared Camera; Martini et al., 2004)
on the Magellan Clay Telescope, at Las Campanas, Chile. PANIC is a
1024$\times$1024 Hawaii infrared array with 0\farcs125 pixels.

The conditions were good to excellent, with seeing between 0\farcs30 and
0\farcs45 (see Table \ref{obs}). We also obtained further imaging in
the I-band using MagIC (the Raymond and Beverly Sackler Magellan
Imaging Camera\footnote{see {\tt
    http://occult.mit.edu/instrumentation/magic/}}; Schectman \&
Johns, 2003), but neither this nor the Y-band were sufficiently deep
to detect Stars A and B and are not considered further. (For
completeness, we note that for the I band, where we have photometric
callibration, the 95\% confidence detection limit is $I>25.1$).

We reduced the images in a standard way, by first subtracting off a
dark frame from each raw image, flat-fielding using  the
median of the images, and then registering and combining them.
For the H- and J-bands, we select the better of each of the
two final images (from the $6^{th}$ and $7^{th}$, respectively) for
analysis rather than combine the images from both nights, since 
the inclusion of the slightly poorer images leads to at best a
marginal improvement in the signal-to-noise ratios. The final JHK$_S$
images we use are shown in Figure \ref{pics}.

We carried out PSF-fitting photometry on each stacked image using {\tt
  DAOPHOT} (Stetson, 1987), using isolated sources on the image to model
the PSF. To calibrate the photometric zero points, we imaged standard
stars P576-F, S165-E, S264-D and S279-F (Persson at al. 1998), took
photometry in a large aperture containing most of the flux and
aperture-corrected the science-frame PSFs using aperture photometry on
the PSF stars (after subtraction of neighboring fainter stars). We
estimate the uncertainty in the photometric zero points to be
$\approx$0.025\,mag for each band. For future reference, we give the
photometry and positions for a number of stars in the field in
Appendix A.

To find the astrometric solutions for our images, we identified stars
from the Guide Star Catalog (GSC) 2.2\footnote{Vizier Catalogue I/271} 
on our J-band image. With 24 stars, the RMS residuals in the solution were 
$\approx0\farcs15$ in each co-ordinate. The systematic uncertainty in
the GSC astrometry is 0\farcs2 to 0\farcs4, so our total uncertainty
in transforming the {\em Chandra} position reported by Israel at
al. (2003) (accurate to 0\farcs7) is 0\farcs8 at 90\%
confidence. This positional error circle is shown in Figure
\ref{pics}. We found astrometric solutions for the rest of our images
by tying them to the Magellan J-band image, which introduces
negligible additional uncertainties. Stars A and B are the two brightest
sources within the error circle. Their positions are ($17^{\mathrm
  d}08^{\mathrm m}46\fs890$, -40\arcdeg08\arcmin52\farcs53) for Star A
and ($17^{\mathrm d}08^{\mathrm m}46\fs904$,
-40\arcdeg08\arcmin52\farcs64) for Star B. (Here the relative
positions are accurate to the digits given, as measured from the NACO
K-band image below, and are on the same astrometric system as the
positions listed in Appendix A from Magellan.)

The separation between Star A and Star B is only $\approx$0\farcs26
(measured from the NACO images below), and so they can only be
measured individually in the K-band, where there is the most flux and
the seeing was best. Even so, one might expect that the
magnitude of the fainter Star B is poorly measured.  Star A's
magnitude, however, should be robust and not affected by the proximity
of Star B. For the H and J bands, Star B could not be measured, but
this should not affect the magnitudes for A too much since the AO
images show that Star B is
over a magnitude fainter in both bands.  The final magnitudes are shown
in Table \ref{mags}. 

\subsection{CFHT}
We retrieved from the CADC archive and re-analyzed the CFHT images
presented by Israel et al. (2003). The AOB uses a wavefront sensor and
deformable mirror to correct for atmospheric distortions, as measured
from a natural guide star. The corrected beam is sent through to the
KIR detector, a Hawaii array with 0\farcs035 pixels (see Rigaut et
al. 1996).
The final reduced stacked images were created
in a standard manner as above; the result is shown in Figure \ref{pics}.

The region of interest is about 12\farcs5 off-axis with respect to the
(fairly faint) AO guide-star and the field is never below an airmass of 1.9 from
CFHT. The AO correction is therefore far from optimal, and the
isoplanatic patch is smaller than the field of view. This means that
the PSF varies from something core-dominated
near the guide star (ideally an Airy pattern) to something more
Gaussian at the furthest point. At no place on the image does the PSF
fit a simple analytic model. This makes photometry difficult, whether
by PSF fitting or by integrating in fixed-sized apertures.

In order to photometer Stars A and B on the CFHT images,
we constructed a PSF based on the average of many stars across the
field with a Lorentzian analytic portion, which hopefully will be a
reasonable fit in the field center. Even though this fit will not be
particularly good, Stars A and B will share the same true PSF (being
so close together), and their relative magnitudes will be accurate. 
Note that although {\tt DAOPHOT} does have the ability to handle a PSF
which varies across the image, there were not enough PSF stars
available on a relatively small field of view for this to work. 

In order to calibrate our magnitudes, we calculated the magnitude
offset relative to the calibrated Magellan images (above) for a number
of isolated stars near to and roughly circularly distributed around
the area of interest. Although the magnitudes show a fair amount of
scatter ($\sigma\sim0.05$\,mag in each band), with some systematic
trends with position, the average offset is well determined and should
be suitable for calibrating the two stars of interest. We estimate a
total uncertainty in the photometric zero points of 0.03\,mag in each
band. See Table \ref{mags} for the final calibrated magnitudes.

Although the K$_S$-band and the K' band do not exactly
overlap, they are close enough that an error due to this is 
negligible compared to the uncertainty in the photometric zero point
above. We see no significant trend in the zero-point offset with $H-K$
color. 

\subsection{VLT/NACO}
The source was imaged in three bands using NAOS-CONICA, the Nasmyth
Adaptive Optics System and Near-Infrared Camera on
VLT Unit telescope 4 (NACO: see Lenzen et al. 2003; Rousset et
al. 2002). CONICA is a 1024 pixel square Aladdin detector (with
0\farcs027 pixel scale).

We retrieved these data from the ESO archive and reduced them in a
similar way to that above. The signal-to-noise ratio of individual
stars is much better than for the CFHT observations (see Figure
\ref{pics}). The isoplanatic 
patch is once more smaller than each of the images (particularly for
the J-band), so we compare the (instrumental) magnitude offsets a set of
isolated, well-measured stars near Stars A and B to those measured
with Magellan. The final calibrated
magnitudes and errors are shown in Table \ref{mags}. We stress
once more that while these magnitudes include systematic uncertainty
from the calibration of the magnitude zero-points, the relative
magnitudes between Stars A and B within an image are very well
determined. 

\begin{deluxetable}{lccccccc}
\tablecaption{List of detections. \label{mags}}
\tabletypesize{\footnotesize}
\tablewidth{0pt}
\tablehead{
\colhead{Observation} & \colhead{$J_A$} & \colhead{$H_A$} &
\colhead{$K_A$} & \colhead{$J_B$} & \colhead{$H_B$} & \colhead{$K_B$}
& \colhead{$\Delta K_{AB}$}}
\startdata
Magellan/PANIC & 20.83(10) & 18.75(5) & 17.45(4) & \nodata & \nodata &
19.26(16)\tablenotemark{b} & 1.81(16) \\
CFHT/AOB & \nodata & 18.82(6) & 17.52(5)\tablenotemark{a} & \nodata & 20.29(13) &
19.02(8)\tablenotemark{a} & 1.50(6)\\
VLT/NACO & 20.88(9) & 18.75(6) & 17.52(3) & 21.89(14) & 20.19(7) &
18.86(3) & 1.344(15)
\enddata
\tablecomments{Numbers in parentheses indicate $1\sigma$ errors in the
  last digit}
\tablenotetext{a}{The K' band is very close to the more common K$_S$
  band.}
\tablenotetext{b}{This measurement is likely affected by the
  proximity of Star~A}
\end{deluxetable}

\section{Results}
We find that the observations above present three items of evidence
that the real counterpart to 1RXS~J170849.0$-$400910 is Star B rather
than Star A as previously reported by Israel et al (2003). These items
may not be conclusive 
individually,  but together make, we argue, a compelling
case. They are discussed separately below. The only argument which
favors Star A is that it lies closer to the center of the positional
error circle, but both lie within the 90\% confidence radius.

Before discussing our lines of argument, we should mention that
the magnitudes we found for Star B are in disagreement
with those presented by Israel et al. (2003), but our magnitudes for
Star A are in good agreement (especially with the NTT data; note that in the
published paper, there was a typo: the magnitude of Star A from the
NTT should have been $17.3\pm0.1$; Israel, 2005 pers. comm.). We suspect
that the discrepancy is due to the use of the on-axis PSF for measuring the
stars (the authors claim a 0\farcs12 FWHM, but this is not the case at
12\farcs5 from the guide star). With our procedure of using stars close
to the sources to create the PSF, and with the much better
signal-to-noise ratio with the NACO images, we believe our photometry highly 
accurate, particularly for the relative magnitudes of Stars A and B.

In Appendix A we also list photometry for faint sources in or near the
positional error circle, which are detected only in the VLT/NACO
K-band image. Without color information or previous measurements, it is not
possible to judge the likelyhood of one of these being the AXP
counterpart, except that they would imply a very large X-ray to
infrared flux ratio in comparison to other AXPs. Given the arguments
in favour of Star B being the counterpart below, we do not consider
these faint sources further, but list their measurements in the
Appendix for completeness. 

\subsection{Variability}
The relative magnitudes ($\Delta K$) given in Table \ref{mags} are
independent of the photometric calibration performed, and show that one of the
two stars has varied (at 3-$\sigma$ significance).

From the magnitudes of the individual stars, it would appear that Star
A shows no significant variability in any band, whereas Star B
apparently brightened. The NACO K-band magnitude is inconsistent at
the 2$\sigma$ level with that from CFHT, and at the 1.9$\sigma$ level
with that from Magellan. A slight brightening is also seen in the
H-band, but this is not statistically significant. Together, it seems
highly likely that Star B has varied.

As a check, the K-band magnitudes of Star 2 (see Figure \ref{photfig})
from Magellan, CFHT and VLT are 18.74(8), 18.88(8) and 18.75(3)
respectively. This shows that this field star is consistent with a
constant brightness, and that the uncertainties in the magnitudes are
reasonable. 

Variability, especially in the K-band is a generic property of AXPs
(Israel et al. 2002; see also Hulleman et al, 2004, Tam et al., 2004,
Durant \& van Kerkwijk, 2005a), so this hint of variability in Star B
and not in Star A is a point in favor of Star B being the true counterpart.

\subsection{Stellar colors}
\begin{figure}
\includegraphics[width=\hsize]{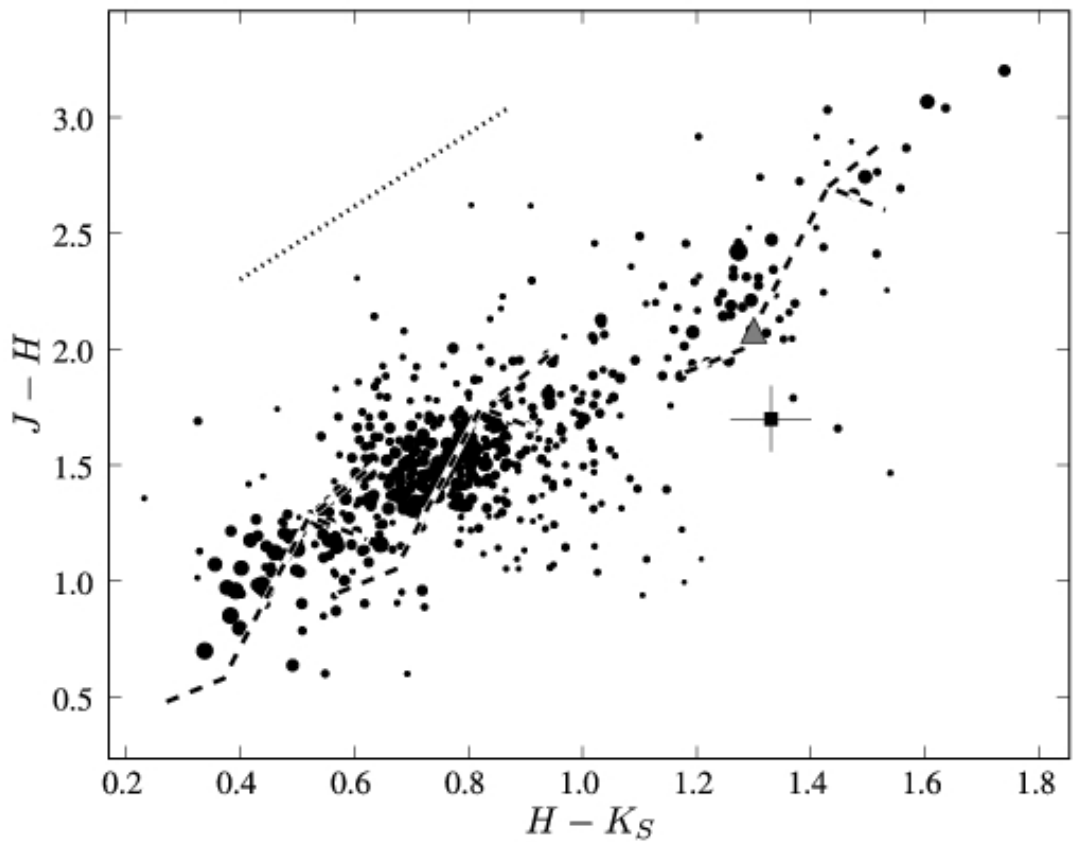}
\caption[0]{Color-color magnitude diagramme for 450 stars detected in
  all three Magellan bands. Symbol sizes are scaled inversely to the
  uncertainties. Stars A and B are shown as the
triangle and filled square with error bars respectively.
 Also shown are the reddening vector for the AXPs estimated reddening,
$A_V=7.4$ (dotted line, e.g. Schlegel et al., 1998) and colors
 expected for main sequence and red giant stars (dashed lines from
 Bessel \& Brett, 1988) reddened by 
$A_V=5.8$, 10.5 and 20, increasing left to right; the upper branches are main
sequences and the lower branches giants.}\label{CCD}
\end{figure}

Figure \ref{CCD} shows all the stars in the Magellan images on a
color-color diagram (after a cut on the $\chi$ goodness-of-fit
diagnostic to reject the worst measured $\sim$10\% of stars). Star A
has been plotted using its Magellan photometry, and Star B using its
VLT/NACO magnitudes relative to Star A (since Star A has not
been seen to vary, this should be secure).

Three different groups of stars with $A_V\approx5.8$, 10.5 and 20 can
be seen in Figure \ref{CCD}. Main sequence and red giant sequences are
shown for these values of reddening. The first two groups are expected from our
analysis of the run of reddening with distance in this direction
(Durant \& van Kerkwijk, 2006b). Star A appears to inhabit the most
reddened group of stars ($A_V\sim20$), with a distance
$d>5$\,kpc. Star B is, however, unusual: less than 5\% of field stars
(about 20 out of 450 in the 40\arcsec\ square region of analysis) are
as far from the expected stellar sequences. Star B does not fit
stellar colors at any reddening.

Other AXPs show non-stellar colors (e.g. Hulleman et al.,
2004) and similar colours have been seen in the infrared for
1E 1048.1$-$5937 ($J-H=0.9(4)$, $H-K_S=1.4(4)$; Wang \& Chakrabarti,
2002) as well as 4U 0142+61 ($J-H=1.2(2)$, $H-K'=1.1(2)$; Israel et
al. 2004). It is unlikely for an ordinary star to occupy the same
region of parameter space as Star B. Its position is consistent with
an infrared excess, possibly from dust emission, as has been found for
4U~0142+61 by Wang et al. (2006). The probability of a chance
coincidence of a star with such colors in the positional error circle
is small ($\approx20 \times \pi 0\farcs8^2 / (40\arcsec)^2\approx$2.5\%).

The color-color diagram hence offers further support for
Star B being the infrared counterpart to 1RXS~J170849.0$-$400910.

\subsection{X-ray to infrared flux ratio}
For the four other AXPs with infrared counterparts, we found that they
were remarkably similar in their X-ray (2--10\,keV) to K-band flux when
not in outburst (Durant \& van Kerkwijk, 2005a):
 all have $F_X/F_K=2700\ldots6000$ (1E~1048.1$-$5937, 4U~0142+61, 1E~2259+589
and XTE~J1810$-$197).

If Star A were the counterpart to 1RXS~J170849.0$-$400910, this would
imply a flux ratio 
$F_X/F_K=660$, whereas for Star B we get (for the range of  magnitudes) 
$F_X/F_K=2300\ldots2870$. (Here we used $F_X=6.4\times10^{-11}$\,erg
s$^{-1}$cm$^{-2}$ [Woods \& Thompson, 2004]  and
$N_H=1.3\times10^{22}$\,cm$^{-2}$ [Durant \& van Kerkwijk,
2006a]). Only the latter value is in the same
range as the other AXPs, which suggests Star B is the more
likely infrared counterpart.

\section{Conclusions}
We presented three lines of evidence that
suggest that Star B is the true counterpart to 1RXS~J170849.0$-$400910:
its variability, its unusual stellar colors and the 
consistency with other AXPs for its inferred X-ray to infrared flux ratio. Of
these, the colors are perhaps the strongest piece of evidence. Together 
they strongly support the identification of Star B as the counterpart.

Despite their proximity on the sky, it seems unlikely that Stars A and
B are physically associated. The hydrogen column for
1RXS~J170849.0$-$400910 of $1.3\times10^{22}$\,cm$^{-2}$ is equivalent
to a visual extinction $A_V\approx7.3$ (Durant \& van
Kerkwijk, 2006a), which is hard to
reconcile with the reddening of $A_V\sim$20 for Star A from its colors. 

With this addition, all but one of the AXPs now have securely
identified optical/infrared counterparts (with the exception being
1E~1841$-$045; see Durant, 2005). Intriguingly, the sources appear to
be rather similar in some of their properties: they have similar X-ray
luminosities (Durant \& van Kerkwijk, 2006b) and similar K-band to
X-ray flux ratios. From the three sources with optical counterparts
however, there appear to be different X-ray to optical flux
ratios. While compared to 4U~0142+61, 1E~1048.1$-$5937 has a similar
X-ray to I-band flux ratio, CXOU~J010043.1$-$721134 has a
significantly lower X-ray to V-band flux ratio (Durant \& van
Kerkwijk, 2005b).  This suggests that the optical and infrared
emission are produced by different mechanisms, with the infrared more
closely tied to the X-ray.  The idea of Wang et al. (2006) of infrared emission
coming from a passively illuminated dusty fall-back disc at the
sublimation radius would seem to be consistent with these data.

\medskip\noindent{\bf Acknowledgments:}
This work made use of the CFHT archive hosted by CADC, of the ESO
VLT archive (for programme 71.D-0503) and of the Vizier archive 
service of the CDS. We used astrometric information from the Guide
Star Catalog 2.2 by the Space Telescope Science Institute and
Osservatorio Astronomico di Torino.
We acknowledge financial support by NSERC. 

\clearpage

\appendix
\section{Photometry}
\begin{figure}[b]
\includegraphics[width=\hsize,angle=270]{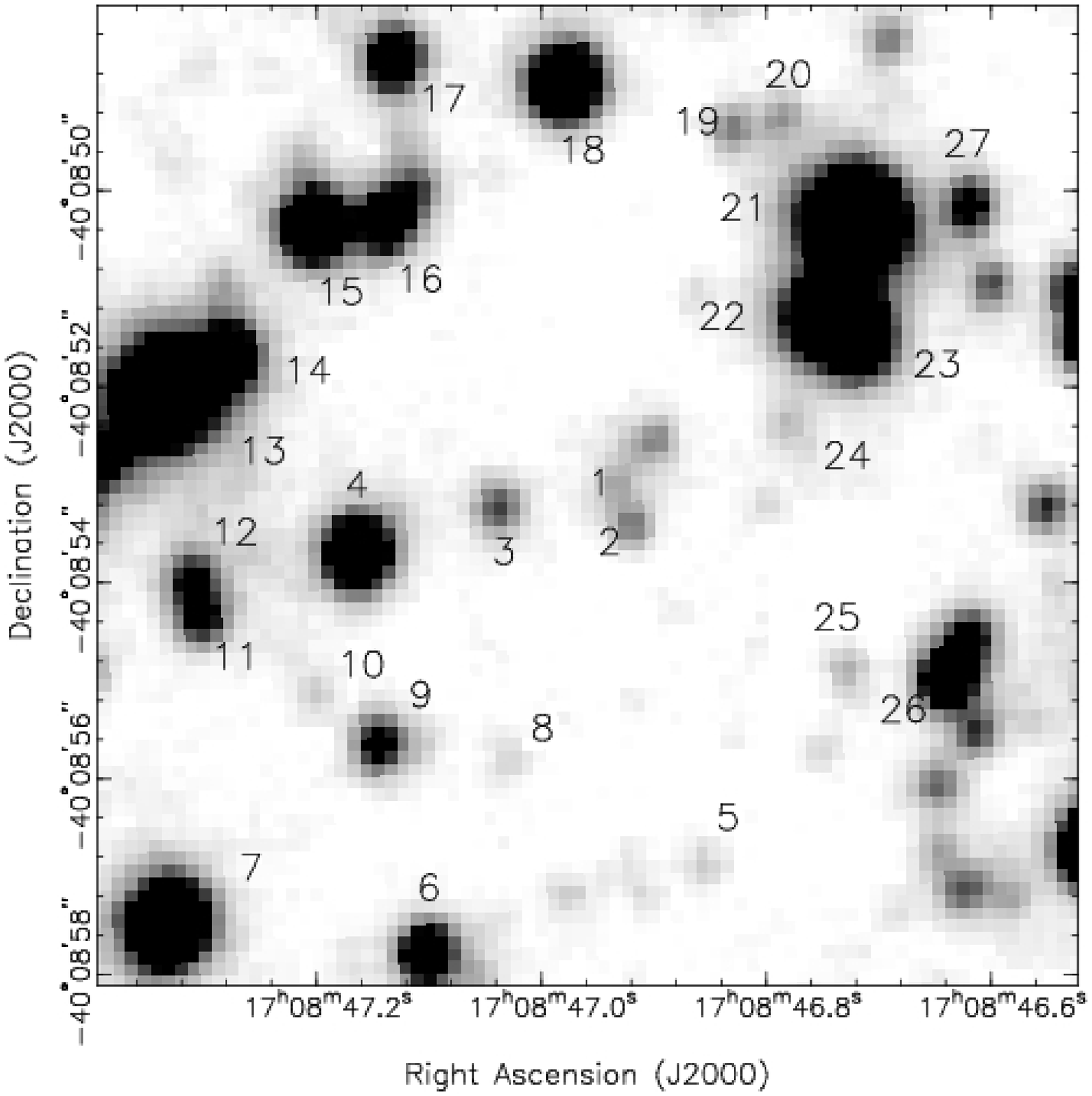}
\caption{Magellan infrared J-band image of the field of
  1RXS~J170849.0$-$400910, with labeled field stars. The stars'
magnitudes and positions are listed in Table \ref{phottab}.
\label{photfig}}
\end{figure}

\begin{figure}[b]
\includegraphics[width=\hsize,angle=270]{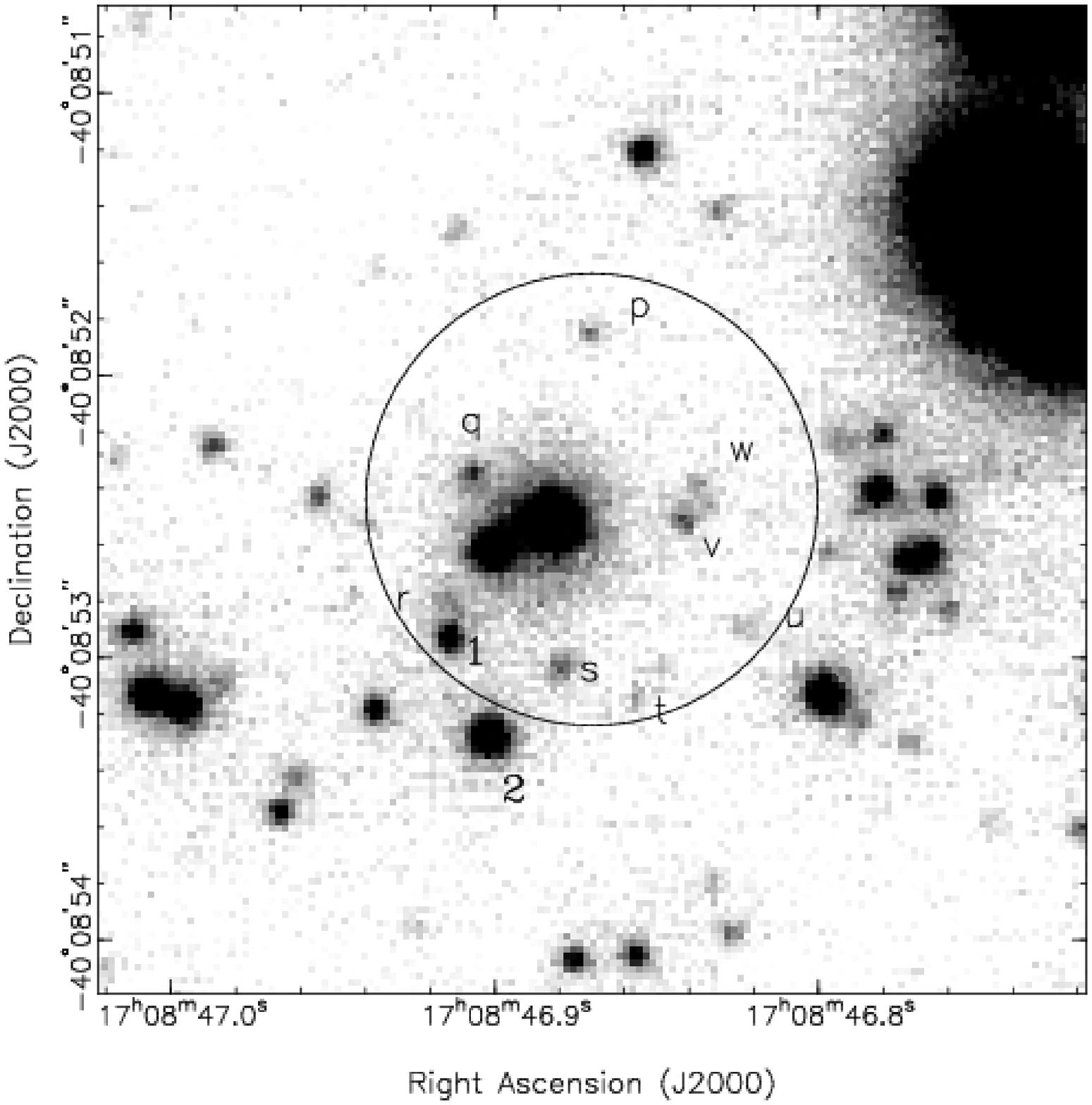}
\caption{VLT/NACO infrared K$_{\mbox{$\scriptstyle{S}$}}-$band image
  of the field of 1RXS~J170849.0$-$400910. Stars within the {\sl Chandra}
  positional error circle are labeled, and their
magnitudes are listed in Table \ref{phottab2}. Also shown are the two
closest numbered stars from Figure \ref{photfig}.
\label{photfig2}}
\end{figure}

In Table \ref{phottab} we list the Magellan JHK magnitudes and
positions of stars in the field, as labeled in Figure
\ref{photfig}. In Table \ref{phottab2} we list the K-band magnitudes
and positions of faint sources detected in or around the position
error circle, as measured in the VLT/NACO observations (Figure
\ref{photfig2}). None of these sources were detected in H or J, with
limits $H>22.1$, $J>22.6$ respectively, at 95\% confidence.

\begin{deluxetable}{clllll}
\tablecaption{Photometry of field stars. \label{phottab}}
\tabletypesize{\footnotesize}
\tablewidth{0pt}
\tablehead{
\colhead{ID\tablenotemark{a}} & \colhead{R.A. (J2000)} & \colhead{dec
  (J2000)} & \colhead{$J$} & \colhead{$H$} & \colhead{$K_S$}}
\startdata
1 & 17:08:46.919&  -40:08:52.97& 21.28(18)& 20.6(2)\tablenotemark{b} &
20.02(6)\tablenotemark{b} \\
2 & 17:08:46.912&  -40:08:53.41& 20.81(11)& 19.34(9)& 18.74(7) \\
3 & 17:08:47.029&  -40:08:53.24& 20.38(7)& 19.30(9)& 18.48(5) \\
4 & 17:08:47.153&  -40:08:53.69& 18.067(10)& 17.216(12)& 16.833(13) \\
5 & 17:08:46.846&  -40:08:56.89& 21.8(3)& 19.53(11)& 18.33(5) \\
6 & 17:08:47.088&  -40:08:57.79& 18.98(2)& 16.352(6)& 15.005(3) \\
7 & 17:08:47.322&  -40:08:57.48& 17.201(5)& 15.601(3)& 14.793(3) \\
8 & 17:08:47.022&  -40:08:55.82& 22.0(4)& 20.4(3)& 19.63(17) \\
9 & 17:08:47.131&  -40:08:55.65& 19.73(4)& 18.48(4)& 17.92(3)\\
10& 17:08:47.190&  -40:08:55.12& 21.9(3)& 20.3(3)& 19.35(14)\\
11& 17:08:47.292&  -40:08:54.35& 19.82(4)& 18.48(4)& 17.80(3)\\
12& 17:08:47.300&  -40:08:53.96& 19.85(4)& 18.43(4)& 17.72(3)\\
13& 17:08:47.329&  -40:08:52.07& 16.316(2)& 15.471(3)& 14.980(3)\\
14& 17:08:47.270&  -40:08:51.70& 18.293(12)& 16.036(5)& 14.838(3)\\
15& 17:08:47.190&  -40:08:50.35& 18.077(10)& 16.997(10)& 16.445(9)\\
16& 17:08:47.124&  -40:08:50.28& 18.850(19)& 17.512(15)& 16.85(13)\\
17& 17:08:47.124&  -40:08:48.65& 18.867(19)& 18.23(3)& 17.74(3)\\
18& 17:08:46.970&  -40:08:48.88& 17.829(8)& 15.814(4)& 14.620(3)\\
19& 17:08:46.816&  -40:08:49.36& 20.84(11)& 19.50(7)& 18.72(7)\\
20& 17:08:46.772&  -40:08:49.27& 21.11(14)& 19.62(11)& 18.88(9)\\
21& 17:08:46.714&  -40:08:50.30& 16.189(2)& 15.520(3)& 15.223(3)\\
22& 17:08:46.728&  -40:08:51.27& 16.745(4)& 15.990(4)& 15.624(3)\\
23& 17:08:46.699&  -40:08:51.59& 19.01(2)& 16.592(7)& 15.319(3)\\
24& 17:08:46.765&  -40:08:52.44& 21.5(2)& 19.82(16)& 18.68(7)\\
25& 17:08:46.714&  -40:08:54.90& 21.5(2)& 19.87(17)& 18.85(9)\\
26& 17:08:46.626&  -40:08:54.98& 18.829(18)& 17.719(19)& 17.241(19)\\
27& 17:08:46.611&  -40:08:50.16& 19.73(4)& 18.23(3)& 17.52(2)
\enddata
\tablecomments{Numbers in parentheses indicate $1\sigma$ errors in the
  last digit, and do not include photometric zero-point uncertainties 
  (approximately 0.025\,mag in each band). Positions are in hours for
  R. A. and degrees for declination. All data
  from Magellan imaging. For Stars A and B see Table~\ref{mags}}
\tablenotetext{a}{As shown in Figure \ref{photfig}}
\tablenotetext{b}{Measured from the NACO images, because of the
  proximity of other sources.}
\end{deluxetable}

\begin{deluxetable}{lccl}
\tablecaption{Photometry and positions of stars in or near the positional error
  circle. \label{phottab2}} 
\tabletypesize{\footnotesize}
\tablewidth{0pt}
\tablehead{
\colhead{ID\tablenotemark{a}} & \colhead{R.A. (J2000)} & \colhead{dec
  (J2000)} & \colhead{$K_S$}}
\startdata
p & 17:08:46.875 & -40:08:51.75 & 21.12(13)\\
q & 17:08:46.912 & -40:08:52.33 & 20.93(11)\\
r & 17:08:46.926 & -40:08:52.85 & 21.6(2)\\
s & 17:08:46.882 & -40:08:53.11 & 20.76(9)\\
t & 17:08:46.853 & -40:08:53.25 & 21.40(17)\\
u & 17:08:46.816 & -40:08:52.96 & 21.6(2)\\
v & 17:08:46.838 & -40:08:52.52 & 20.85(10)\\
w & 17:08:46.838 & -40:08:52.38 & 21.36(17)\\
\enddata
\tablecomments{Numbers in parentheses indicate $1\sigma$ errors in the
  last digit, and do not include photometric zero-point uncertainties 
  (approximately 0.03\,mag).}
\tablenotetext{a}{As shown in Figure \ref{photfig2}}
\end{deluxetable}

\end{document}